\documentstyle[aps,12pt,epsf,amsfonts,amssymb,amsbsy]{revtex}

\newcommand\VV{\setbox0=\hbox{V}\hbox{\rm V\raise\ht0
  \hbox to0pt{\hss\vbox to0pt{\hbox{v}\vss}}}}
\def\slashchar#1{\setbox0=\hbox{$#1$}           
   \dimen0=\wd0                                 
   \setbox1=\hbox{/} \dimen1=\wd1               
   \ifdim\dimen0>\dimen1                        
      \rlap{\hbox to \dimen0{\hfil/\hfil}}      
      #1                                        
   \else                                        
      \rlap{\hbox to \dimen1{\hfil$#1$\hfil}}   
      /                                         
   \fi}                                         %

\begin{document}

\begin{center}
{\large \bf Doubly heavy baryons\footnote{We omit the indices ${QQ^\prime}$,
when they actually do not lead to misidentificaion with usual strange baryons.}
$\Omega_{QQ^\prime}$ vs. $\Xi_{QQ^\prime}$ in sum rules of NRQCD
}\\
\vspace*{5mm}
V.V. Kiselev, \\[1mm]
{\it State Research Center of Russia 
"Institute for High Energy Physics",\\ Protvino, Moscow region, 
142284 Russia}\\[2mm]
and\\[1mm]
A.E. Kovalsky\\[1mm]
{\it Moscow Institute of Physics and Technology,\\ 
Dolgoprudny, Moscow region, 141700 Russia}
\end{center}

\begin{abstract}
In the framework of two-point sum rules of NRQCD, the masses and couplings of
doubly heavy baryons to the corresponding quark currents are evaluated with
account of coulomb-like corrections in the system of doubly heavy diquark as
well as the contribution of nonperturbative terms determined by the quark,
gluon, mixed condensates and the product of gluon and quark condensates. The
introduction of nonzero light quark mass destroys the factorization of baryon
and diquark correlators even at the perturbative level and provides the better
convergency of sum rules. We estimate the difference $M_{\Omega} - M_{\Xi} =
100\pm10~$MeV. The ratio of baryonic constants $|Z_{\Omega}|^2/|Z_{\Xi}|^2$ is
equal to $1.3\pm0.2$ indicating the violation of SU(3) flavor symmetry for the
doubly heavy baryons.
\end{abstract}

\vspace*{5mm}
PACS numbers: 12.39.Hg, 14.20.Lq, 14.20.Mr

\section{Introduction}

Testing the QCD dynamics of heavy quarks in various conditions provides us with
a qualitative and quantitative knowledge that allows us to distinguish fine
complex effects caused by the electroweak nature of CP-violation or physics
beyond the Standard Model. The list of hadrons containing the heavy quarks as
available to the experimental observations and measurements, was recently
enriched by a new member, the long-lived $B_c$-meson, in addition to the heavy
quarkonia $\bar b b$ and $\bar c c$ as well as the mesons and baryons with a
single heavy quark. This success of CDF Collaboration in the first observation
of $B_c$-meson \cite{CDF} was based on the progress of experimental technique
in the reconstruction of rare processes with heavy quarks by use of vertex
detectors. This experience supports a hope to observe other rare long-lived
doubly heavy hadrons. i.e. the baryons containing two heavy quarks. As expected
they have production rate \cite{Production} and lifetime \cite{Lifetime}
similar to the $B_c$-meson ones.  

In the present paper we investigate the two-point sum rules \cite{SVZ} of NRQCD
\cite{NRQCD}. The light quark-doubly heavy diquark structure of baryon leads to
the definite expressions for baryonic currents written in terms of
nonrelativistic heavy quark fields. To relate the nonrelativstic heavy quark
correlators to the full QCD ones we need to take into account the hard gluon
corrections by means of solving the renormalization group equation known up to
the two-loop accuracy. 
 
The convergency of sum rules results are essentially improved by account for a
nonzero light quark mass. As was mentioned in \cite{KO} the sum rules stability
can be achieved by destroying the baryon-diquark factorization in the
correlators. The convergency was obtained due to taking into account the
nonperturbative interactions caused by higher dimension operators in contrast
to ref.\cite{old}, where a signigicant instability of results were observed in
full QCD sum rules with no product of quark and gluon condensates. We show that
for the strange $\Omega_{QQ^\prime}$ baryons this factorization is broken
already in the perturbative limit which allows us to introduce a new criterion
for the determination of baryon masses since we observe the stability of sum
rules for the masses obtained from both correlators standing in front of two
independent Lorentz structures for the spinor field of $\Omega_{QQ^\prime}$.

Moreover our choice of baryonic current is convenient to take into account
the $\alpha_s/v$ coulomb-like corrections \cite{Coulomb} inside the doubly
heavy diquark.

In Section II we describe the scheme of calculation. There we define the
currents and represent the spectral densities in the NRQCD sum rules for
various operators included into the consideration. The numerical estimates in
comparison with the values obtained in potential models are given in Section
III. The results are summarized in Conclusion.

\section{NRQCD sum rules for doubly heavy baryons}
\subsection{Description of the method}
In order to determine the masses and coupling constants of baryons in sum
rules we consider the two point correlators of interpolating baryon currents.
The quantum numbers of doubly heavy diquark in the ground states are given by
its spin and parity, so that $j^P_d=1^+$ or $j^P_d=0^+$ (if the identical heavy
quarks form the diquark then the scalar state $j^P_d=0^+$ is forbidden). Adding
the light quark to form the baryon, we obtain the pair of degenerate states
$j^P=\frac{1}{2}^+$ and $j^P=\frac{3}{2}^+$ for the baryons\footnote{The
superscript $\diamond$ denotes various electric charges depending on the flavor
of the light quark.} $\Xi_{cc}^{\diamond}$, $\Xi_{bc}^{\diamond}$,
$\Xi_{bb}^{\diamond}$ and $\Xi_{cc}^{*\diamond}$, $\Xi_{bc}^{*\diamond}$,
$\Xi_{bb}^{*\diamond}$ with the vector diquark, and
$j^P=\frac{1}{2}^+$ for the $\Xi^{\prime \diamond}_{bc}$ baryons with the
scalar diquark. Unlike the case of baryons with a single heavy quark \cite{HB},
there is the only independent current component for each ground state. We find
 \begin{eqnarray}
 J_{\Xi^{\prime \diamond}_{QQ^{\prime}}} &=& [Q^{iT}C\tau\gamma_5
 Q^{j\prime}]q^k\varepsilon_{ijk},\nonumber\\
 J_{\Xi_{QQ}^{\diamond}} &=& [Q^{iT}C\tau\boldsymbol{\gamma}^m
 Q^j]\cdot\boldsymbol{\gamma}_m\gamma_5
 q^k\varepsilon_{ijk},
 \label{def}\\
 J_{\Xi_{QQ}^{*\diamond}}^n &=& [Q^{iT}C\tau\boldsymbol{\gamma}^n
 Q^j]q^k\varepsilon_{ijk}+\frac{1}{3}\boldsymbol{\gamma}^n
 [Q^{iT}C\boldsymbol{\gamma}^m
 Q^j]\cdot\boldsymbol{\gamma}_m q^k\varepsilon_{ijk},\nonumber
 \end{eqnarray}
 where $J_{\Xi_{QQ}^{*\diamond}}^n$ satisfies the spin-3/2 condition
 $\boldsymbol{\gamma}_n J_{\Xi_{QQ}^{*\diamond}}^n = 0$. The flavor matrix
 $\tau$ is antisymmetric for $\Xi^{\prime \diamond}_{bc}$ and symmetric for
 $\Xi_{QQ}^{\diamond}$ and $\Xi_{QQ}^{*\diamond}$. Here $T$ means
 transposition, $C$ is the charge conjugation matrix.
 
The matrix structure of correlator for two baryonic currents with the spin of
$1/2$ has the form
 \begin{equation}
 \Pi(w)=i \int d^4 x e^{i p x} \langle 0|T{J(x), \bar J(0)}|0
 \rangle=\slashchar v F_1(w)+F_2(w), 
 \end{equation}
where $w$ is defined by $p^2=({\cal{M}}+w)^2$, ${\cal{M}}=m_Q+m_{Q'}+m_s$,
$m_{Q,Q'}$ are the heavy quark masses and $m_s$ is the strange quark mass. The
appropriate definitions of scalar formfactors for the 3/2-spin baryon are given
by the following:
 \begin{equation}
 \Pi_{\mu\nu}(w) = i\int d^4x e^{ipx}\langle 0|T\{J_\mu (x),\bar
 J_\nu (0)\}|0\rangle = -g_{\mu\nu}[\slashchar{v} \tilde F_1(w) +
 \tilde F_2(w)]+\ldots,\label{2vcor}
 \end{equation}
where we do not concern for distinct Lorentz structures. The scalar correlators
$F$ can be evaluated in a deep euclidean region by employing the Operator
Product Expansion (OPE) in the framework of NRQCD,
 \begin{equation}
 F_{1,2}(w) = \sum_d C^{(1,2)}_d(w)O_d,
 \end{equation}
where $O_d$ denotes the local operator with a given dimension $d$:
 $O_0 = \hat 1$, $O_3 = \langle\bar qq\rangle$, $O_4 =
 \langle\frac{\alpha_s}{\pi}G^2\rangle$, \ldots,
and the functions $C_d(w)$ are the corresponding Wilson coefficients of OPE.
For the contribution of quark condensate operator we explore the following OPE
up to the terms of the fourth order in $x~$ (the derivation is presented in
Appendix):
 \begin{eqnarray}
 \langle 0|T{s_i^a(x)\bar s_j^b(0)}|0
 \rangle&=&-\frac{1}{12}\delta^{ab}\delta_{ij}\langle \bar s s\rangle\cdot
 \nonumber\\
 &&\left[1+\frac{x^2(m_0^2-2
 m_s^2)}{16}+\frac{x^4(\pi^2\langle
 \frac{a_s}{\pi}G^2\rangle-\frac{3}{2}m_s^2(m_0^2-m_s^2))}{288}\right]\nonumber
 \\
 &+& i m_s \delta^{ab}  x_{\mu}\,\gamma^{\mu}_{ij}\cdot\langle \bar q
 q\rangle\cdot\left[\frac{1}{48}+\frac{x^2}{24^2}\left(\frac{3
 m_0^2}{4}-m_s^2\right)\right].
 \label{Exp}
 \end{eqnarray}
Note that at $m_s\neq0$ the expansion of quark condensate gives contributions
in both correlators in contrast with the sum rules for $\Xi_{QQ^\prime}$
\cite{KO}, where putting $m_s = 0$ and neglecting the higher condensates, the
authors found the factorization of diquark correlator in $F_2$ and full
baryonic correlator in $F_1$. This fact was the physical reason for the
divergency of SR method. 
 
We write down the Wilson coefficient in front of unity and quark-gluon
operators by making use of the dispersion relation over $w$,
 \begin{equation}
 C_d(w)=\frac{1}{\pi}\int_0^{\infty}\frac{\rho_d(\omega)
 d\omega}{\omega-w},
 \end{equation}
where $\rho$ denotes the imaginary part in the physical region of NRQCD. 
 
To relate the NRQCD correlators to the real hadrons, we use the dispersion
representation for the two point function with the physical spectral density
given by the appropriate resonance and continuum part. The coupling constants
of baryons are defined by the following expressions:
\begin{eqnarray}
\langle 0| J(x)|{\Xi(\Omega)_{QQ}^{\diamond}}(p)\rangle & = & i 
Z_{\Xi(\Omega)_{QQ}^{\diamond}} u(v,M_{\Xi(\Omega)}) e^{ip x}, \nonumber
\\
\langle 0| J^m(x)|{\Xi(\Omega)_{QQ}^{*\diamond}}(p,\lambda)\rangle & = &
i Z_{\Xi(\Omega)_{QQ}^{*\diamond}} u^m(v,M_{\Xi(\Omega)}) e^{ip x}, \nonumber 
\end{eqnarray}
where the spinor field with the four-velocity $v$ and mass $M$ (the mass of
baryon) satisfies the equation $\slashchar{v} u(v,M) = u(v,M)$, and 
$u^m(v,M)$ denotes the transversal spinor. 

Then we use the nonrelativistic expressions for the physical spectral
functions
\begin{equation}
\rho^{phys}_{1,2}(\omega)=\frac{M}{2{\cal{M}}} |Z|^2
\delta(\bar \Lambda-\omega),
\end{equation}
where we have performed the substitution $\delta(p^2-M^2)\to\frac{1}{2
{\cal{M}}} \delta(\bar \Lambda-w)$, here $\bar \Lambda$ is the binding energy
of baryon and $M={\cal{M}}+\bar \Lambda$. The nonrelativistic dispersion
relation for the hadronic part of sum rules has the form
\begin{equation}
\int \frac{\rho^{phys}_{1,2} d\omega}{\omega-w}=\frac{1}{2{\cal{M}}}
\frac{|Z|^2
}{\bar \Lambda-w}.
\end{equation}

We suppose that the continuum densities starting from the threshold
$\omega_{cont}$, is modelled by the NRQCD expressions. Then, in the sum rules
equalizing the correlators calculated in NRQCD and given by the physical
states, the integration above $\omega_{cont}$ cancel each other in two sides of
sum rules. Further, we write down the correlators in the deep underthreshold
point of $w=-{\cal{M}}+t$ with $t\to0$. 
 
Now the sum rules in the scheme of moments with respect to $t$ can be written
down as follows:
\begin{equation}
\frac{1}{\pi}\int_0^{\omega_{cont}} \frac{\rho_{1,2}\,
d\omega}{(\omega+{\cal{M}})^n}=\frac{M }{2{\cal{M}}}\frac{|Z|^2}{M^n},
\end{equation}
where $\rho_j$ contains the contributions given by various operators in OPE for
the corresponding scalar functions $F_j$.
Introducing the following notation for the $n$-th moment of two point 
correlation function:
\begin{equation}
{\cal M}_n	=\frac{1}{\pi}\int_0^{\omega_{cont}}\frac{\rho (\omega
)\,d\omega}
{(\omega +{\cal{M}})^{n+1}},
\end{equation}
we can obtain the estimates of baryon mass $M$, for example, as the
following:
\begin{equation}
M[n] = \frac{{\cal M}_n}{{\cal M}_{n + 1}},
\end{equation}
and the coupling is determined by the expression
\begin{equation}
|Z[n]|^2 =\frac{{2 \cal{M}}}{ M} {\cal M}_n M^{n + 1},
\end{equation}
where we see the dependence of sum rule results on the scheme parameter.

\subsection{Calculating the spectral densities}

In this subsection we present analytical expression for the perturbative
spectral functions in the NRQCD approximation. The evaluation of spectral
densities involves the standard use of Cutkosky rules \cite{Cutkovsky}, with
some modification motivated by NRQCD 
 \begin{eqnarray}
 {\rm heavy\; quark:}\;\;&& \frac{1}{p_0-(m+\frac{{\bf p}^2}{2m})}\to 2\pi
 i\cdot\delta (p_0-(m+\frac{{\bf p}^2}{2m})), \nonumber\\
 {\rm light\; quark:}\;\;&& \frac{1}{p^2-m^2} \to 2\pi i\cdot\delta
 (p^2-m^2).\nonumber
 \end{eqnarray}
We derive the spin symmetry relations for all the  spectral densities due to
the fact that in the leading order of the heavy quark effective theory the
spins of heavy quarks are decoupled, so 
\begin{equation}
\rho_{1,\Omega_{QQ}^{\diamond}}=3 \rho_{1,\Omega_{QQ'}^{'\diamond}}=3
\rho_{1,\Omega_{QQ}^{*\diamond}},
\end{equation}
\begin{equation}
\rho_{2,\Omega_{QQ}^{\diamond}}=3 \rho_{2,\Omega_{QQ'}^{'\diamond}}=3
\rho_{2,\Omega_{QQ}^{*\diamond}},
\end{equation}
and we have the following relation for the baryon couplings in NRQCD:
\begin{equation}
|Z_{\Omega}|^2=3|Z_{\Omega'}|^2=3|Z_{\Omega^*}|^2.
\end{equation}
Using the smallness of the strange quark mass we use the following expansions
in $m_s$ for the perturbative spectral densities standing in front of unity
operator ($m_{QQ'}=m_Q m_{Q'}/(m_Q+m_{Q'})$ is the reduced diquark mass,
${\cal{M}}_{diq}=m_Q+m_{Q'}$):
\begin{equation}
\rho_{1,\Omega^{'\diamond}_{QQ'}}(\omega)=\frac{\sqrt{2} (m_{QQ'}
\omega)^{3/2} }{15015 \pi^3
({\cal {M}}_{diq}+\omega)^3}(\eta_{1,0}(\omega)+m_s
\eta_{1,1}(\omega)+m_s^2 \eta_{1,2}(\omega)),
\end{equation}
where we have found
\begin{eqnarray}
\eta_{1,0}(\omega) &=&16\omega^2(429 {\cal {M}}_{diq}^3+715 {\cal {M}}_{diq}^2
\omega+403{\cal {M}}_{diq} \omega^2+77\omega^3),\nonumber\\
\eta_{1,1}(\omega)&=& 104  \omega (231 {\cal {M}}_{diq}^3+297 {\cal
{M}}_{diq}^2
\omega+121 {\cal {M}}_{diq}
\omega^2+15
\omega^3),\nonumber\\
\eta_{1,2}(\omega)&=&\frac{10 }{({\cal {M}}_{diq}+\omega)^2}(3003
{\cal {M}}_{diq}^5+9009 {\cal {M}}_{diq}^4
\omega+9438
{\cal {M}}_{diq}^3 \omega^2\nonumber\\
&&+4290
{\cal {M}}_{diq}^2 \omega^3+871 {\cal {M}}_{diq} \omega^4 +77 \omega^5).
\end{eqnarray}
The first term of this expansion reproduces the result obtained in \cite{KO}.
A new feature is the appearance of nonzero perturbative
$\rho_{2,\Omega^{'\diamond}_{QQ'}}$ density which is proportional to $m_s$,
\begin{equation}
\rho_{2,\Omega^{'\diamond}_{QQ'}}(\omega)=\frac{2 \sqrt{2} \omega (m_{QQ'}
\omega)^{3/2} m_s}{105 \pi^3
({\cal {M}}_{diq}+\omega)^2} (\eta_{2,0}+m_s \eta_{2,1}+m_s^2 \eta_{2,2}),
\end{equation}
and
\begin{eqnarray}
\eta_{2,0} &=& 42 \omega ({\cal {M}}_{diq}^2+48 {\cal {M}}_{diq} \omega +14
\omega^2),\nonumber\\
\eta_{2,1} &=& 3(35 {\cal {M}}_{diq}^2+28 {\cal {M}}_{diq}
\omega+5\omega^2),\nonumber\\
\eta_{2,2} &=& \frac{1}{({\cal {M}}_{diq}+\omega)^2} (105
{\cal{M}}_{diq}^3+315{\cal {M}}_{diq}^2
\omega+279 {\cal {M}}_{diq} \omega^2+77 \omega^3).
\end{eqnarray}

The account for the coulomb-like interaction leads to the finite
renormalization of the diquark spectral densities before the integration over
the diquark invariant mass by the introduction of Sommerfeld factor {\bf {C}},
so that
\begin{equation}
\rho_{diquark}^{\bf{C}}=\rho_{diquark}^{bare} \cdot {\bf{C}}
\end{equation}
with
\begin{equation}
{\bf C} = \frac{2\pi\alpha_s}{3v_{QQ'}}\left [ 1- \exp\left
(-\frac{2\pi\alpha_s}{3v_{QQ'}} \right)\right ]^{-1},
\end{equation}
where $v_{12}$ denotes the relative velocity of heavy quarks inside the
diquark, and we have taken into account the color anti-triplet structure of
diquark. The relative velocity is given by the following expression:
\begin{equation}
v_{QQ'} = \sqrt{1-\frac{4m_Q m_{Q'}}{Q^2-(m_Q-m_{Q'})^2}},
\end{equation}
where $Q^2$ is the square of heavy diquark four-momentum. In NRQCD we take the
limit of low velocities, so that denoting the diquark invariant mass squared by
$Q^2=({\cal{M}}_{diq}+\epsilon)^2$,  we find
$$
{\bf C} = \frac{2\pi\alpha_s}{3v_{QQ'}},\;\;\;
v_{QQ'}^2 = \frac{\epsilon}{2m_{QQ'}},
$$
at $\epsilon\ll m_{QQ'}$. 

The corrected spectral densities are equal to
\begin{equation}
\rho_{1}^{{\bf C}}(\omega)=\frac{m_{QQ'}^2 \alpha_s \omega (2
{\cal M}_{diq}+\omega)}{6
\pi^2 ({\cal M}_{diq}+\omega)^3}(\eta_{1,0}^{{\bf C}}+m_s\eta_{1,1}^{{\bf
C}}+m_s^2\eta_{1,2}^{{\bf C}}),
\end{equation}
where
\begin{eqnarray}
\eta_{1,0}^{{\bf C}}&=&(2 {\cal M}_{diq}+\omega)^2\omega^2,\nonumber\\
\eta_{1,1}^{{\bf C}}&=&\frac{3(2 {\cal M}_{diq}+\omega)\omega}{({\cal
M}_{diq}+\omega)}(4 {\cal M}_{diq}^3+6{\cal M}_{diq}^2 \omega+4 {\cal M}_{diq}
\omega^2+\omega^3),\nonumber\\
\eta_{1,2}^{{\bf C}}&=&\frac{1}{({\cal M}_{diq}+\omega)^2}(12 {\cal
M}_{diq}^4+24 {\cal M}_{diq}^3
\omega+32 {\cal M}_{diq}^2 \omega^2+20 {\cal M}_{diq} \omega^3+5 \omega^4).
\end{eqnarray}
We see that the first term again reproduces the result of \cite{KO}.
For the $\rho_{2,\Omega^{'\diamond}_{QQ'}}^{\bf {C}}$ we find
\begin{equation}
\rho_{2}^{\bf {C}}=\frac{m_s m_{QQ'}^2 (2 {\cal M}_{diq}+\omega)\omega \alpha_s
}{2 \pi({\cal M}_{diq}+\omega)^2}(\eta_{2,0}^{{\bf C}}+m_s \eta_{2,1}^{{\bf
C}}+m_s^2 \eta_{2,2}^{{\bf C}}),
\end{equation}
\begin{eqnarray}
\eta_{2,0}^{{\bf C}}&=&(2 {\cal M}_{diq}+\omega)\omega ,\nonumber\\
\eta_{2,1}^{{\bf C}}&=& \frac{2}{{\cal M}_{diq}+\omega}(2 {\cal M}_{diq}^2+2
{\cal M}_{diq}
\omega+\omega^2), \nonumber\\
\eta_{2,0}^{{\bf C}}&=&\frac{2}{({\cal M}_{diq}+\omega)^2}(2 {\cal M}_{diq}^2+2
{\cal M}_{diq} \omega+\omega^2).
\end{eqnarray}
The use of these expansions numerically leads to very small deviations from the
exact integral representations of spectral densities (about 0.5\%), but they
are more convenient in calculations. 

The contribution to the moments of the spectral densities determined by the
light quark condensate can be calculated by the exploration of (\ref{Exp})
\begin{eqnarray}
{\cal M}_{\bar q q}^{(1)}(n)&=&-\frac{(n+1)!}{n!} {\cal{P}}_1 {\cal
M}^{diq}(n+1)+\frac{(n+3)!}{n!} {\cal{P}}_3 {\cal M}^{diq}(n+3)\nonumber\\
{\cal {\cal M}}_{\bar q q}^{(2)}(n)&=&{\cal{P}}_0 {\cal
M}^{diq}(n)-\frac{(n+2)!}{n!} {\cal{P}}_2 {\cal M}^{diq}(n+2)+\frac{(n+4)!}{n!}
{\cal{P}}_4 {\cal M}^{diq}(n+4),
\end{eqnarray}
where we have introduced the coefficients of expansion in $x$ by ${\cal{P}}_i$
(see (\ref{Exp}) and Appendix). The n-th moment of two-point correlator
function of diquark is denoted by ${\cal M}^{diq}(n)$. Then the diquark
spectral density takes the following form:
\begin{equation}
\rho_{diq}=\frac{12 \sqrt 2 m_{QQ'}^{3/2} \sqrt\omega}{\pi}, 
\end{equation}
which has to be multiplied by the Sommerfeld factor ${\bf{C}}$, wherein the
variable $\epsilon$ is substituted by $\omega$, since in this case there is no
integration over the quark-diquark invariant mass. This corrected density is
\begin{equation}
\rho_{diq}^{{\bf{C}}}=\frac{48 \pi \alpha_s m_{QQ'}^2}{3},
\end{equation}
and it is independent of $\omega$.

The corrections due to the gluon condensate are given by the density
\begin{equation}
\rho_{1}^{G^2}(\omega)=\frac{(m_Q^2+m_{Q'}^2+11 m_Q m_{Q'}) m_{QQ'}^{5/2}\sqrt
\omega}{21\cdot2^{10} \sqrt 2 \pi 
m_Q^2
m_{Q'}^2 ({\cal{M}}_{diq}+\omega)^2} (\eta^{G^2}_{1,0}+m_s
\eta^{G^2}_{1,1}+m_s^2
\eta^{G^2}_{1,2}),
\end{equation}
with
\begin{eqnarray}
\eta^{G^2}_{1,0}&=& 84 {\cal{M}}_{diq}^3+196 {\cal{M}}_{diq}^2
\omega+133{\cal{M}}_{diq} \omega^2+11 \omega^3,\nonumber\\
\eta^{G^2}_{1,1}&=& -\frac{2(210{\cal{M}}_{diq}^3+70 {\cal{M}}_{diq}^2
\omega+21{\cal{M}}_{diq}
\omega^2+3 \omega^3)}{{\cal{M}}_{diq}+\omega},\nonumber\\
\eta^{G^2}_{1,2}&=& \frac{2(210{\cal{M}}_{diq}^3+70 {\cal{M}}_{diq}^2
\omega+21{\cal{M}}_{diq}
\omega^2+3 \omega^3)}{({\cal{M}}_{diq}+\omega)^2},
\end{eqnarray}
where we again make the expansion in $m_s$. In the case of nonzero quark mass
we get the nonzero density proportional to $m_s$,
\begin{equation}
\rho_{2}^{G^2}(\omega)=\frac{m_s (m_Q^2+m_{Q'}^2+11 m_Q m_{Q'})
m_{QQ'}^{5/2}\sqrt
\omega}{3\cdot2^{9} \sqrt 2 \pi 
m_Q^2
m_{Q'}^2 ({\cal{M}}_{diq}+\omega)} (\eta^{G^2}_{2,0}+m_s
\eta^{G^2}_{2,1}),
\end{equation}
with 
\begin{eqnarray}
\eta^{G^2}_{2,0} &=& -(9 {\cal{M}}_{diq}+\omega),\nonumber\\
\eta^{G^2}_{2,1} &=& \frac{9 {\cal{M}}_{diq}+\omega}{{\cal{M}}_{diq}+\omega}.
\end{eqnarray}
For the product of condensates $\langle \bar q q \rangle \langle
\frac{\alpha_s}{\pi} G^2\rangle$, wherein the gluon fields are connected to the
heavy quarks, it is convenient to compute the contribution to the two-point
correlation function itself. We have found
\begin{equation}
F^{\bar q q G^2}_2(\omega)=-\frac{m_{QQ'}^{5/2}(m_Q^2+m_{Q'}^2+11 m_Q
m_{Q'})}{ 2^{9} \sqrt 2 \pi  m_{Q}
m_{Q'}(-\omega)^{5/2}},
\end{equation}
and we put $F^{\bar q q G^2}_1(\omega)=0$, since we have restricted the
dimension of condensate operators, while the corresponding term in $F_1$ would
appear in the fifth order in expansion (\ref{Exp}). The result is given in the
form, which allows the analytical continuation over $\omega=-{\cal{M}}+w$.

\subsection{Matching with full QCD} 
To connect the NRQCD sum rules to the baryonic couplings in full QCD we have to
use the relation
$$
J^{QCD} = {\cal K}_J(\alpha_s,\mu_{\rm soft},\mu_{\rm hard}) \cdot J^{NRQCD},
$$
where the coefficient ${\cal K}_J(\alpha_s,\mu_{\rm soft},\mu_{\rm hard})$
depends on
the soft normalization scale $\mu_{\rm soft}$. The ${\cal K}$-factor obeys the
matching condition at the hard scale $\mu_{\rm hard}={\cal{M}}_{diq}$ and is
determined by the anomalous dimensions of effective baryonic currents which are
independent of the diquark spin in the leading order. They are known up to the
two loop accuracy \cite{anom}. In our consideration we use the one-loop
accuracy, since the subleading corrections in the first $\alpha_s$ order are
not available yet. So,
\begin{eqnarray}
\gamma &=& \frac{d\ln {\cal K}_J(\alpha_s,\mu_{\rm soft},\mu_{\rm hard})}{d\ln
(\mu)} =
\sum_{m=1}^\infty\left(\frac{\alpha_s}{4\pi}\right)^m\gamma^{(m)},
	   \nonumber\\
\gamma^{(1)} &=& \Big(-2C_B(3a-3)+3C_F(a-2)\Big), \label{gamma}
\end{eqnarray}
where $C_F=(N_c^2-1)/2N_c$, $C_B=(N_c+1)/2N_c$ for $N_c=3$ and $a$ is the gauge
parameter. In Feynman gauge $a=1$, and we get $\gamma^{(1)}=-4$. So, in the
leading logarithmic approximation and to the one-loop accuracy we find
\begin{equation}
{\cal K}_J(\alpha_s,\mu_{\rm soft},\mu_{\rm hard}) = \left(
\frac{\alpha_s(\mu_{\rm hard})}{\alpha_s(\mu_{\rm
soft})}\right)^{\frac{\gamma^{(1)}}{2
\beta_0}},
\end{equation}
where $\beta_0=11-2/3 N_F=9$. Further, we determine the soft normalization
scale for the NRQCD estimates by the average momentum transfer inside the
doubly heavy diquark, so that $\mu_{\rm soft}^2=2 m_{QQ'}T_{diq}$, where
$T_{diq}$
is the kinetic energy in the system of two heavy quarks, which is
phenomenologically independent of the heavy quark flavours and approximately
equal to 0.2 GeV \cite{RQ}. Then, the coefficients ${\cal K}_J$ are equal to
\begin{equation}
{\cal K}_{\Omega_{cc}} \approx 1.95,\;\;\;
{\cal K}_{\Omega_{bc}} \approx 1.52,\;\;\;
{\cal K}_{\Omega_{bb}} \approx 1.30,
\end{equation}
with the characteristic uncertainty about $10\%$ basically due to the variation
of hard and soft scale points $\mu_{\rm hard,soft}$. Note that the values of
${\cal K}_J$ do not change the estimates of baryon masses, but they are
essential in the evaluation of baryon couplings.

\section{Numerical results}
Evaluating the two-point sum rules, we explore the scheme of moments. We point
out the well-known fact that an essential part of uncertainties is caused by
the variation of heavy quark masses. Indeed, the results of sum rules for the
systems containing two heavy quarks strongly depend on the choice of masses,
and this fact allows us to pin down the values of masses with a high precision
up to 20 MeV \cite{3pt}, so that $m_b=4.60\pm0.02~$GeV, $m_c=1.40\pm0.03~$GeV,
which are extracted from the two-point sum rules for the families of $\Upsilon$
and $\psi$. To get these values we have use the correlators evaluated up to the
same accuracy in $\alpha_s$, i.e. we have put the quark loop with the
appropriate Sommerfeld factor. Then, the stability criterion for the leptonic
constant of heavy quarkonium strictly fixes the heavy quark masses, which are
close to the results of \cite{benhomel}. The same sum rules are also explored
to estimate the couplings determining the coulomb-like interactions inside the
heavy quarkonia
\begin{equation}
\alpha_s(b\bar b) = 0.37,\;\;\;
\alpha_s(c\bar b) = 0.45,\;\;\;
\alpha_s(c\bar c) = 0.60,\;\;\;
\end{equation}
since they fix the absolute normalization of corresponding leptonic constants
known experimentally.

Since the squared size of diquark is two times larger than that of the meson
the effective coulumb constants have to be rescaled according to the equation
of evolution in QCD. We use the one-loop evolution equation
$$
\alpha_s(QQ')=\frac{\alpha_s(Q\bar Q')}{1-\frac{9 }{4 \pi}\alpha_s(Q\bar Q')
\ln 2}.
$$
So,
\begin{equation}
\alpha_s(b b) = 0.45,\;\;\;
\alpha_s(b c) = 0.58,\;\;\;
\alpha_s(c c) = 0.85.\;\;\;
\end{equation}
The values of condensates are taken in the ranges $\langle \bar q q
\rangle=-(0.26\div 0.27~\mbox{GeV})^3$, $m_0^2 = 0.75\div 0.85\; \mbox{GeV}^2$,
$\langle \frac{\alpha_s}{\pi} G^2\rangle=(1.7\div
1.8)\cdot10^{-2}~\mbox{GeV}^4$.
 
The main source of uncertainties in the ratios of the baryonic couplings is the
ratio of the condensates of the strange quark and light quark. We use ${\langle
\bar s s \rangle}/{\langle \bar q q\rangle }=0.8\pm0.2$ that corresponds to the
variations of the sum $(m_u+m_d)[1\, {\rm GeV}] = 12\div14~$ MeV \cite{Pivss}.
 
\begin{figure}[th]
\begin{center}
\begin{picture}(100,200)
\put(-100,0){\epsfxsize=10cm \epsfbox{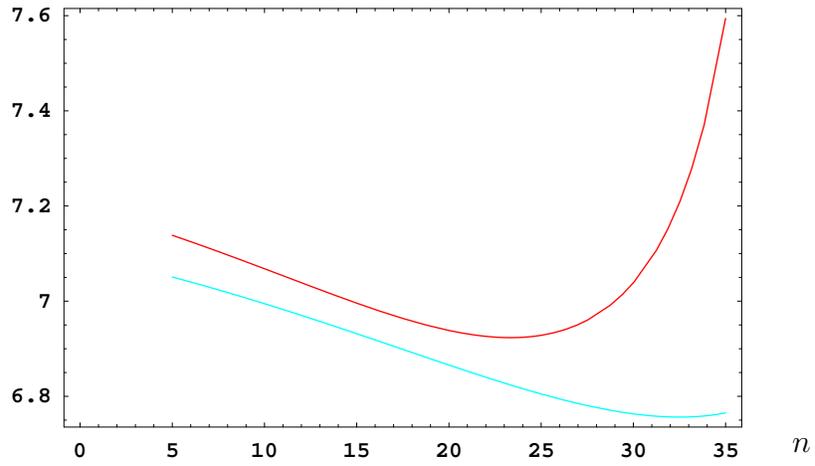}}
\put(200,5){$n$}
\put(-110,200){$M_{\Xi,\Omega_{bc}}$, GeV}
\end{picture}
\end{center}
\caption{The $\Xi_{bc}$(lower curve) and $\Omega_{bc}~$(upper curve) masses
obtained in the NRQCD sum rules from the first correlator $F_1$.}
\label{M1}
\end{figure}

\begin{figure}[ph]
\begin{center}
\begin{picture}(100,200)
\put(-100,0){\epsfxsize=10cm \epsfbox{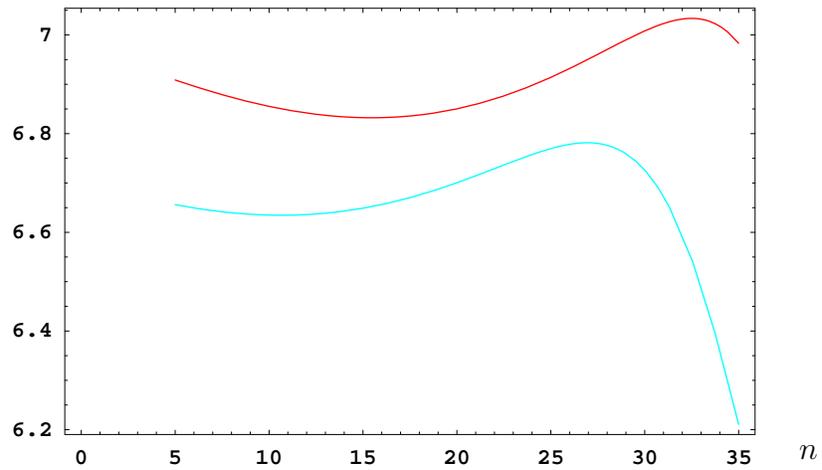}}
\put(200,5){$n$}
\put(-110,200){$M_{\Xi,\Omega_{bc}}$, GeV}
\end{picture}
\end{center}
\caption{The $\Xi_{bc}$(lower curve) and $\Omega_{bc}~$(upper curve) masses
obtained in the NRQCD two point sum rules from the second correlator $F_2$.}
\label{M2}
\end{figure}

\begin{figure}[ph]
\begin{center}
\begin{picture}(100,200)
\put(-100,0){\epsfxsize=10cm \epsfbox{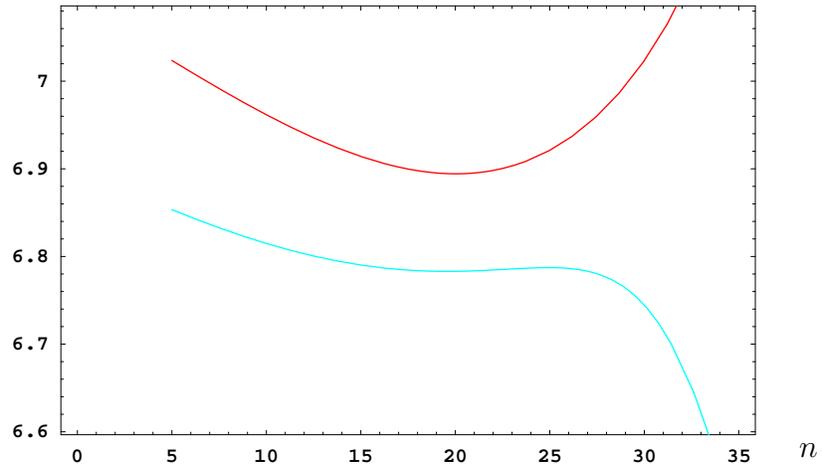}}
\put(200,5){$n$}
\put(-110,200){$M_{\Xi,\Omega_{bc}}$, GeV}
\end{picture}
\end{center}
\caption{The $\Xi_{bc}$(lower curve) and $\Omega_{bc}~$(upper curve) masses
obtained by averaging the results from both correlators.}
\label{M}
\end{figure}

So, we have described the set of parameters entering the scheme of
calculations. In Figs. \ref{M1}-\ref{M} we present the results of the two-point
sum rules for the masses of $\Xi_{bc}$ and $\Omega_{bc}$~(the figures for the
other baryons are similar). For the $\Omega_{bc}$-baryons one can observe the
stability of mass with respect to the changing of the moment numbers in both
correlators. We suppose it is connected with the destroying of
diquark-$\Omega~$baryon factorization in the perturbative limit in contrast to
the $\Xi$-baryons. The stability regions for $F_1$ and $F_2$ are not coincide
because the contributions of higher dimension operators become valuable at the
different numbers of moments. However, the quantity $1/2(M_1[n]+M_2[n])$ has
the larger stability region, and we explore this fact to determine the $\Omega$
baryons masses as well as that of $\Xi$ baryons. The theoretical uncertainties
in the $\Omega$-baryon masses are mainly determined by the difference between
the values of baryon masses at the regions of stability. 

\begin{figure}[th]
\begin{center}
\begin{picture}(100,200)
\put(-100,0){\epsfxsize=10cm \epsfbox{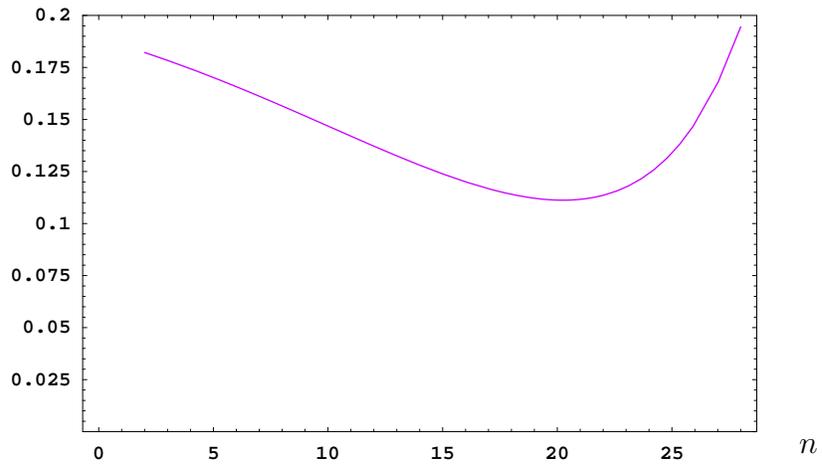}}
\put(200,5){$n$}
\put(-110,200){$\Delta M$, GeV}
\end{picture}
\end{center}
\caption{The mass difference $\Delta M=M_{\Omega_{bc}}-M_{\Xi_{bc}}$ obtained
from the results shown in Fig \ref{M}.}
\label{Mres}
\end{figure}

Then, we investigate the difference between the masses $1/2((M_{1,\Omega} +
M_{2,\Omega}) - (M_{1,\Xi}+M_{2,\Xi})) $ shown in Fig. \ref{Mres}. In our
scheme of baryon masses determination  this quantity has the meaning of the
difference between the $\Omega$ and $\Xi$ baryon masses. It has the large
region of stability and is determined with a good precision. We obtain
$$
\Delta M=M_{\Omega_{bb}}-M_{\Xi_{bb}} = M_{\Omega_{cc}}-M_{\Xi_{cc}} =
M_{\Omega_{bc}}-M_{\Xi_{bc}}=100\pm10\; {\rm MeV.}
$$

\begin{figure}[ph]
\begin{center}
\begin{picture}(100,200)
\put(-100,0){\epsfxsize=10cm \epsfbox{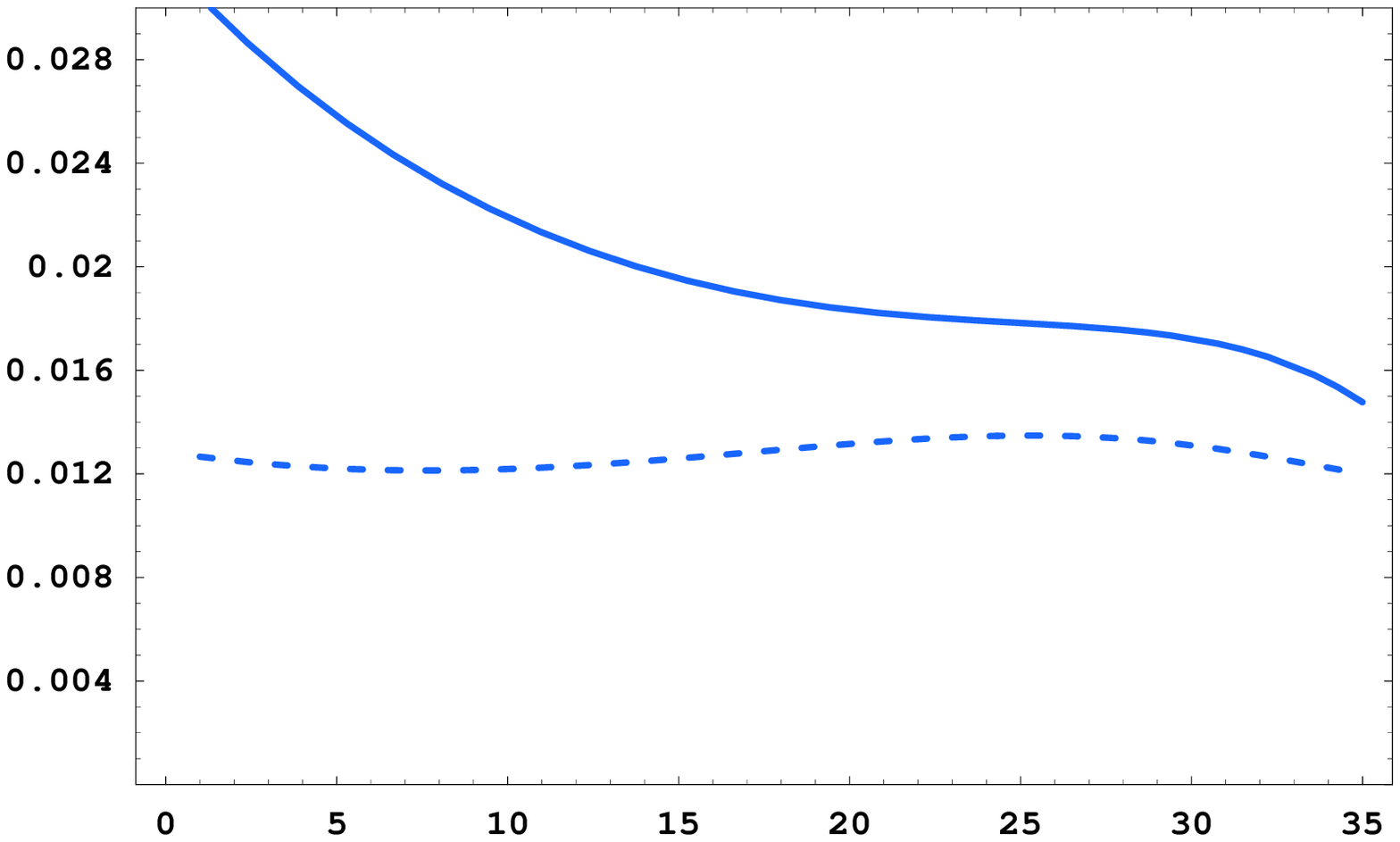}}
\put(200,5){$n$}
\put(-110,200){$|Z_{\Omega_{bc}}|^2$, $\mbox {GeV}^6$}
\end{picture}
\end{center}
\caption{The couplings $|Z_{\Omega_{bc}}^{(1,2)}|^2$ calculated in NRQCD sum
rules for the formfactors $F_1$ and $F_2~$(solid and dashed lines,
correspondingly) in the scheme of moments for the spectral densities.}
\label{Zbcs}
\end{figure}

\begin{figure}[ph]
\begin{center}
\begin{picture}(100,200)
\put(-100,0){\epsfxsize=10cm \epsfbox{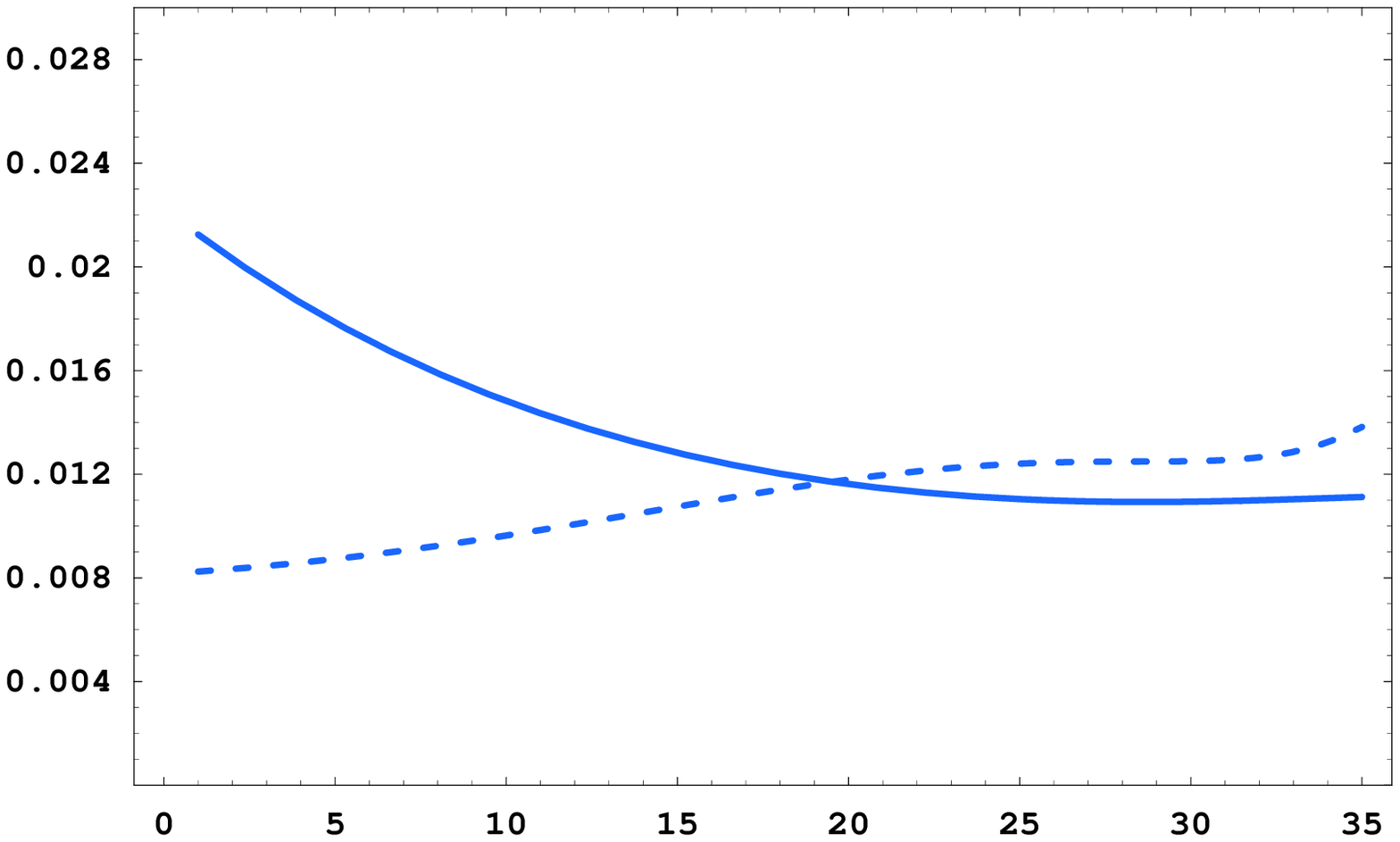}}
\put(200,5){$n$}
\put(-110,200){$|Z^{NR}_{\Xi_{bc}}|^2$, $\mbox {GeV}^6$}
\end{picture}
\end{center}
\caption{The couplings $|Z_{\Xi_{bc}}^{(1,2)}|^2$ calculated in NRQCD sum rules
for the formfactors $F_1$ and $F_2~$(solid and dashed lines, correspondingly)
in the scheme of moments for the spectral densities.}
\label{Zbc}
\end{figure}

The uncertainty in the $\Xi$-baryons masses are determined through the
uncertainty in the $\Omega$-baryons masses and that of in $\Delta M$. So, for
the masses we find the following results:
\begin{equation}
 \begin{array}{lcrrlcrr}
M_{\Omega_{bc}} &=& 6.89\pm0.05 & \mbox{GeV}, &
   M_{\Xi_{bc}} &=& 6.79\pm0.06 & \mbox{GeV},\\
M_{\Omega_{bb}} &=& 10.09\pm0.05 &\mbox{GeV}, &
   M_{\Xi_{bb}} &=& 10.00\pm0.06 &\mbox{GeV},\\
M_{\Omega_{cc}} &=& 3.65\pm0.05  &\mbox{GeV}, &  
   M_{\Xi_{cc}} &=& 3.55\pm0.06  &\mbox{GeV}.
 \end{array}
\end{equation}
The obtained values are in agreement with the calculations in the framework of
nonrelativistic potential models \cite{PM1,PM2}, though the models based on the
calculation of three body systems with the pair interactions \cite{PM2} give
slightly higher values of masses. In \cite{KO} the other method of baryon mass
determination was used, since the quantities $M_{1,\Xi}$ and $M_{2,\Xi}$
separately have no good stability in the sum rules. So, the difference of
$M_1-M_2$ close to zero was stable. The use of $\frac{1}{2}(M_1+M_2)$ stability
criterion results in the $\Xi_{QQ'}$ masses coinciding with those of \cite{KO}
up to 10 MeV. Figs.\ref{Zbcs}, \ref{Zbc} show the dependence of baryon
couplings calculated in the moment scheme of NRQCD sum rules. Numerically, we
find
\begin{equation}
 \begin{array}{lcrrlcrr}
|Z_{\Omega_{cc}}|^2 &=& (10.0\pm1.2)\cdot10^{-3} &\mbox{GeV}^6, &
   |Z_{\Xi_{cc}}|^2 &=& (7.2\pm0.8)\cdot10^{-3}  &\mbox{GeV}^6,\\[1mm]
|Z_{\Omega_{bc}}|^2 &=& (15.6\pm1.6)\cdot10^{-3} &\mbox{GeV}^6, &
   |Z_{\Xi_{bc}}|^2 &=& (11.6\pm1.0)\cdot10^{-3} &\mbox{GeV}^6,\\[1mm]
|Z_{\Omega_{bb}}|^2 &=& (6.0\pm0.8)\cdot10^{-2}  &\mbox{GeV}^6, &
   |Z_{\Xi_{bb}}|^2 &=& (4.2\pm0.6)\cdot10^{-2}  &\mbox{GeV}^6.
\end{array}
\label{nrqcd}
\end{equation}

\begin{figure}[th]
\begin{center}
\begin{picture}(100,200)
\put(-100,0){\epsfxsize=10cm \epsfbox{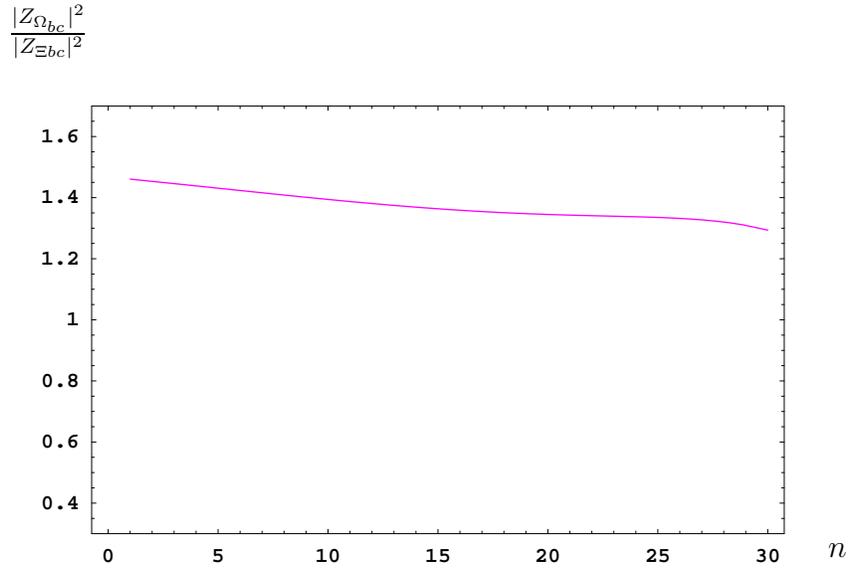}}
\put(200,5){$n$}
\put(-110,200){$\frac{|Z_{\Omega_{bc}}|^2}{|Z_{\Xi{bc}}|^2}$}
\end{picture}
\end{center}
\caption{The ratio $\frac{|Z_{\Omega_{bc}}|^2}{|Z_{\Xi{bc}}|^2}$ calculated in
NRQCD sum rules in the scheme of moments for the spectral densities at
${\langle \bar s s \rangle}/{\langle \bar q q\rangle }=0.8$.}
\label{Ratio}
\end{figure}

In Fig. \ref{Ratio} we present the sum rules results for the ratio of baryonic
constants $|Z_{\Omega_{bc}}|^2/|Z_{\Xi_{bc}}|^2$. We have also found
$$
|Z_{\Omega_{bc}}|^2/|Z_{\Xi_{bc}}|^2=|Z_{\Omega_{cc}}|^2/|Z_{\Xi_{cc}}|^2=
|Z_{\Omega_{bb}}|^2/|Z_{\Xi_{bb}}|^2=1.3\pm0.2.
$$
The uncertainty of this result as was mentioned above is mainly connected with
the pourly known ratio of $\langle \bar s s \rangle /\langle \bar q q
\rangle=0.8\pm0.2$.

For the sake of comparison, we derive the relation between the baryon coupling
and the wave function of doubly heavy baryon evaluated in the framework of
potential model, where we have used the approximation of quark-diquark
factorization. So, we find
\begin{equation}
|Z^{\rm PM}| = 2 \sqrt{3} |\Psi_d(0)\cdot \Psi_{l,\,s}(0)|,
\end{equation}
where $\Psi_d(0)$ and $\Psi_{l,\,s}(0)$ denote the wave functions at the origin
for the doubly heavy diquark and light (strange) quark-diquark systems,
respectively. In the approximation used, the values of $\Psi(0)$ were
calculated in \cite{PM1} in the potential by Buchm\" uller--Tye \cite{BT}, so
that
\begin{eqnarray}
\sqrt{4\pi}\;|\Psi_l(0)|    &=& 0.53\; {\rm GeV}^{3/2},\nonumber\\
\sqrt{4\pi}\;|\Psi_s(0)|    &=& 0.64\; {\rm GeV}^{3/2},\nonumber\\
\sqrt{4\pi}|\Psi_{cc}(0)| &=& 0.53\; {\rm GeV}^{3/2},\nonumber\\
\sqrt{4\pi}|\Psi_{bc}(0)| &=& 0.73\; {\rm GeV}^{3/2},\nonumber\\
\sqrt{4\pi}|\Psi_{bb}(0)| &=& 1.35\; {\rm GeV}^{3/2}.\nonumber
\end{eqnarray}
In the static limit of potential models, these parameters result in the
estimates 
\begin{eqnarray}
|Z^{\rm PM}_{\Omega_{cc}}|^2~ =~ 8.8\cdot 10^{-3}\; {\rm GeV}^6,\;
|Z^{\rm PM}_{\Xi_{cc}}|^2 &=& 6.0\cdot 10^{-3}\; {\rm GeV}^6,~\nonumber\\
|Z^{\rm PM}_{\Omega_{bc}}|^2 ~=~ 1.6\cdot 10^{-2}\; {\rm GeV}^6,\;
|Z^{\rm PM}_{\Xi_{bc}}|^2 &=& 1.1\cdot 10^{-2}\; {\rm GeV}^6,~\label{pot}\\
|Z^{\rm PM}_{\Omega_{bb}}|^2~ =~ 5.6\cdot 10^{-2}\; {\rm GeV}^6,\;
|Z^{\rm PM}_{\Xi_{bb}}|^2 &=& 3.9\cdot 10^{-2}\; {\rm GeV}^6.~\nonumber
\end{eqnarray}
The estimates in the potential model (\ref{pot}) are close to the values
obtained in the sum rules of NRQCD (\ref{nrqcd}). We also see that the
SU(3)-flavor splitting for the baryonic constants $|Z_{\Omega}|^2/|Z_{\Xi}|^2$
is determined by the ratio $|\Psi_s(0)|^2/|\Psi_l(0)|^2=1.45$ which is in
agreement with the sum rules result. The values obtained in the NRQCD sum rules
have to be multiplied by the Wilson coefficients coming from the expansion of
full QCD operators in terms of NRQCD fields, as they have been estimated by use
of corresponding anomalous dimensions. This procedure results in the final
estimates
\begin{eqnarray}
|Z_{\Omega_{cc}}|^2&=&(38\pm5)\cdot10^{-3}~\mbox{GeV}^6,\;\;
|Z_{\Xi_{cc}}|^2=(27\pm3)\cdot10^{-3}~\mbox{GeV}^6, \nonumber\\
|Z_{\Omega_{bc}}|^2&=&(36\pm4)\cdot10^{-3}~\mbox{GeV}^6,\;\;
|Z_{\Xi_{bc}}|^2=(27\pm3)\cdot10^{-3}~\mbox{GeV}^6, \\
|Z_{\Omega_{bb}}|^2&=&(10\pm1)\cdot10^{-2}~\mbox{GeV}^6,\;\;
|Z_{\Xi_{bb}}|^2=(70\pm 8)\cdot10^{-3}~\mbox{GeV}^6. \nonumber
\end{eqnarray}

\section{Conclusion}
 
In this paper the NRQCD	sum rules applied to the doubly heavy baryons have been
considered. The nonrelativistic approximation for the heavy quark fields allows
us to fix the structure of baryonic currents (the light quark-doubly heavy
diquark) and to take into account the coulomb-like interactions inside the
doubly heavy diquark. The presence of both the nonzero mass of light quark and
the contribution of nonperturbative terms of the quark, gluon, mixed
condensates and the product of condensates destroys the factorization of the
correlators. This fact provides the convergency of sum rules for each
correlator and allows us to obtain the reliable results for the masses and
baryonic constants, which agree with the estimates in the framework of
potential models. We also have calculated the mass splitting of $\Omega$ and
$\Xi$ doubly heavy baryons and the ratio of baryonic constants
$|Z_{\Omega}|^2/|Z_{\Xi}|^2$.

The authors are grateful to prof. A.K.Likhoded for stimulating
discussions.

This work is in part supported by the Russian Foundation for Basic Research,
grants 99-02-16558 and 00-15-96645.

\section{Appendix}
Here the derivation of expansion (\ref{Exp}) is briefly presented. The
calculations are done in the technique of fixed point gauge \cite{FPG}, so we
write down the expansion of quark field:
 $$q(x)=q(0)+x^{\alpha} D_{\alpha}q(0)+\frac{1}{2}~ x^{\alpha} x^{\beta}
 D_{\alpha}
D_{\beta}q(0)+...,$$
and in the evaluation of $\langle0|T{q_i^a(x)\bar q^b_j(0)}|0\rangle$, where
$i$ and$j$ are the spinor indices, $a,b$ are the color indices, we have to know
how to get the vacuum average of type $\langle0|D_{\alpha}...D_{\omega}q(0)~
\bar q(0)|0\rangle$. The main formulae are the followings:

the definitions of condensates 
$$
\langle q^a_i(0)
 \bar q^b_j(0)\rangle_0=-\frac{1}{12}\delta^{ab}\delta_{ij}\langle \bar q q
 \rangle,
$$ 
$$
 \langle G_{\alpha \beta}^{a} G_{\alpha' \beta'}^{a'}\rangle=\frac{\delta^{a
 a'}}{96}(
 g_{\alpha \alpha'} g_{\beta \beta'}-g_{\alpha \beta'} g_{\alpha' \beta})
 \langle G^2 \rangle, 
$$
$$
\langle \bar q~ i g
G_{\alpha \beta}^a t^a \sigma_{\alpha \beta} q\rangle_0=m_0^2 \langle \bar q q
\rangle, 
$$

the commutator of covariant derivatives
 $$
 [D_{\alpha},D_{\beta}]=-i g G_{\alpha \beta}^{a} t^a,
 $$ 

 and the equation of  motion for the spinor field 
 $$
\slashchar D q=-i m_q q.
 $$
Form the last two equations we  derive the so-called quadratic Dirac equation,
$$
D^2 q=- m_q^2 q+\frac{\sigma_{\alpha \beta}}{2} i g  G_{\alpha \beta}^a t^a q .
$$ 
Now it is an easy challenge to obtain the first term in expansion~(\ref{Exp}).

Since the tensor $x_{\alpha}...x_{\omega}$ is the symmetric one, we may perform
the symmetrization  
$$
D_{\alpha}...D_{\omega} \to \{D_{\alpha},...,D_{\omega}\}_+ ,
$$ 
to find the n-th term of expansion for $\langle \bar q(x) q(0) \rangle$,  which
equals 
$$
\frac{1}{n!}~x_{\alpha}...x_{\omega}
\langle \bar q(0)~D_{\alpha}...D_{\omega} q(0)
\rangle=\frac{1}{n!}~x_{\alpha}...x_{\omega}
\langle \bar q(0)~\{D_{\alpha},...,D_{\omega}\}_+~ q(0) \rangle.
$$ 
Note, the tensor $\langle \bar q(0)~\{D_{\alpha}...D_{\omega}\}_+~ q(0)
\rangle$ is also symmetric one.
 
The second term of expansion is derived from
$$
\langle \{D_{\alpha}, D_{\beta}\}_+ q_{\rho}^i(0)~ \bar
q_{\eta}^j(0)\rangle=-2! ~{\cal P}_2\cdot g_{\alpha \beta} \delta^{ij}
\delta_{\rho \eta}
\langle \bar q q\rangle,
$$ 
and the coefficient ${\cal P}_2$ is determined by contracting the indices
$\alpha, \beta$ and using the quadratic Dirac equation,
$$
{\cal P}_2=(m_0^2-2m_q^2)/192.
$$

The third term can be derived from the following structure:
$$
\langle \{D_{\alpha}, D_{\beta}, D_{\delta}\}_+ q^{i}_{\rho}(0) \bar
q_{\eta}(0) \rangle=-3!~{\cal P}_3\cdot \delta^{ij} ((\gamma_{\alpha})_{\rho
\eta}
g_{\beta
\delta}+(\gamma_{\beta})_{\rho \eta} g_{\alpha \delta}+(\gamma_{\delta})_{\rho
\eta} g_{\alpha \beta}) \langle \bar q q \rangle. 
$$
Then, contracting $\alpha$ and $\beta$ and using of the equation of
motion, the quadratic Dirac equation and the commutation relation, we obtain
$$
{\cal P}_3=-i~m_q(3m_0^2/4-m_q^2)/576.
$$ 
This includes the evaluation of vacuum
averages 
$$
\langle D^2D_{\alpha}q(0)~\bar q(0)\rangle,~ \langle D_{\beta}
D_{\alpha} D_{\beta} q(0)~ \bar q(0)\rangle~ \mbox{and}~ \langle D_{\alpha} D^2
q(0)~\bar q(0)\rangle.
$$

Considering the structure 
$$
\langle \{D_{\alpha}, D_{\beta}, D_{\delta}, D_{\xi} \}_+ q^{i}_{\rho}(0) \bar
q_{\eta}(0) \rangle=-4!~{\cal P}_4 \cdot \delta^{ij} \delta_{\rho
\eta}(g_{\alpha
\beta}~g_{\delta \xi}+g_{\alpha \delta}~g_{\beta \xi}+g_{\alpha \xi}~g_{\delta
\beta}) \langle
\bar q q\rangle
$$
contracted over any pair of indices, we derive 
$$
{\cal P}_4=(\pi^2\langle\alpha_s/\pi~ G^2\rangle+3/2~m_q^2(m_q^2-m_0^2))/3456.
$$
Here we evaluated the following types
of vacuum expectations: 
$$
\langle D^2 D^2 q(0)~\bar q(0) \rangle,~\langle D_{\alpha} D_{\beta} D_{\alpha}
D_{\beta} q(0)~\bar q(0) \rangle,~ \langle D_{\alpha} D^2 D_{\alpha} q(0)~\bar
q(0) \rangle.
$$

Then the OPE for the quark condensate can be expressed in terms of
${\cal P}_i$ by
$$
\langle q^{i}_{\rho}(x) \bar q^{j}_{\eta}(0)\rangle=-\delta^{ij}\langle \bar q
q
\rangle \cdot ({\cal P}_0 \delta_{\rho \eta}+{\cal P}_1
x_{\alpha}\gamma^{\alpha}_{\rho
\eta}+{\cal P}_2 \delta_{\rho \eta}  x^2+{\cal P}_3
x_{\alpha}\gamma^{\alpha}_{\rho \eta} x^2+ {\cal P}_4
\delta_{\rho \eta}
x^4),
$$

with ${\cal P}_0=1/12$, and ${\cal P}_1=-im_q/48 $.


\newpage

\end{document}